\documentclass[11pt,a4paper,oneside,final,notitlepage,onecolumn]{article}
\usepackage{amsfonts}
\usepackage{epsf}
\usepackage[dvips]{color}

\renewcommand{\baselinestretch}{1.5}

\begin{document}


\newtheorem{theorem}{Theorem}[section]
\newtheorem{lemma}{Lemma}[section]
\newtheorem{proposition}{Proposition}[section]
\newtheorem{corollary}{Corollary}[section]
\newtheorem{conjecture}{Conjecture}[section]
\newtheorem{example}{Example}[section]
\newtheorem{definition}{Definition}[section]
\newtheorem{remark}{Remark}[section]
\newtheorem{exercise}{Exercise}[section]
\newtheorem{axiom}{Axiom}[section]



\author{Istv\'{a}n R\'{a}cz\thanks{%
email: istvan@sunserv.kfki.hu} 
\\ 
KFKI Research Institute for Particle and Nuclear Physics\\
H-1525 Budapest 114 P.O.B. 49, Hungary}

\title{Does the third law of black hole thermodynamics really have a
serious failure?}
\maketitle

\vskip 2cm

\begin{abstract}

The almost perfect correspondence between certain laws of classical
black hole mechanics and the ordinary laws of thermodynamics is
spoiled by the failure of the conventional back hole analogue of the
third law.  Our aim here is to contribute to the associated discussion
by flashing light on some simple facts of black hole physics.
However, no attempt is made to lay to rest the corresponding long
lasting debate. Instead, merely some evidence is provided to make it
clear that although the borderline between extremal and non-extremal
black holes is very thin they are essentially different. Hopefully, a
careful investigation of the related issues will end up with an
appropriate form of the third law and hence with an unblemished
setting of black hole thermodynamics.

\end{abstract}

\vfill\eject

\setcounter{equation}{0}

During the past three decades a considerable development was
established in general relativity centered around black hole physics.
Perhaps the most remarkable of the underlying results is the
manifestation of the strong analogies between certain laws of
classical black hole mechanics and the ordinary laws of thermodynamics
(for a recent review see e.g.  \cite{w3}). A significant enhancement
of this striking mathematical correspondences was established by
Hawking's discovery \cite{h1} that black holes radiate as perfect
black bodies. This provided, in particular, a crucial justification
that surface gravity, $\kappa$, of a stationary black hole is in a
direct connection with the truly physical temperature of the associated
thermal state.

\medskip

However, the nearly perfect correspondence between the relevant basic
laws of black hole mechanics and ordinary thermodynamics
seems to be spoiled by the explicit failure of the conventional black
hole analogue of the third law. The strong version of the third law asserts
that the entropy, $S$, of a system tends to a universal constant,
which may be taken to be zero, while its temperature, $T$, approaches
absolute zero \cite{hu}. The failure of this law in black hole
mechanics is usually demonstrated by recalling the entropy and
temperature expressions relevant for Kerr-Newman black holes. As it is
well-known the entropy of such a black hole can be given as
\begin{equation}
S=A/4=\pi \left[r_+^2 + a^2\right],\label{e1}
\end{equation} 
while the associated temperature reads as
\begin{equation}
T=\kappa/2\pi= \frac {1}{2\pi} \frac {r_+-m} {r_+^2 + a^2},\label{t1}
\end{equation} 
where
\begin{equation}
r_+=m+\sqrt{m^2-a^2-q^2}
\end{equation}
moreover, $m$, $a$ and $q$ denote, respectively, the mass, the angular
momentum to mass ratio, $J/m$, and the electric charge of the black
hole. (Throughout we use units such that $G=c=\hbar=k=1$.) Clearly,
by setting the above parameters so that $m^2=a^2+q^2$ the associated
temperature vanishes. On the other hand, the non-vanishing entropy
expression $\pi[m^2 + a^2]$ depends on the state parameters hence  
it cannot be 
equal to a ``universal constant''. Thereby the existence of these
extreme Kerr-Newman black holes is considered to demonstrate the
violation of the black hole analogue of the strong version of the
third law of thermodynamics.

Although seemingly sensible classical and quantum statistical systems
can be prepared so that the above strong version of the third law of
thermodynamics is violated (see e.g. \cite{w1}), the validity
of this law has been experimentally verified for all substances so far
investigated.  Therefore one would expect that it may  hold for any
``reasonable thermodynamical system''.  Hence, the failure of the
third law in black hole mechanics indicates that there has to be
something ``exotic'' about the thermodynamic properties of extremal
black holes.

\medskip

In classical thermodynamics it follows from the strong form of the
third law that a system cannot be cooled to absolute zero by finite
change of its thermodynamical parameters \cite{hu}. The last statement
is referred to sometimes as the ``weak form of the third law'' or as
`Nernst theorem'. In fact an analogous unattainability of absolute
zero, i.e. that of an extreme black hole state, can be justified by
explicit calculations (see e.g. \cite{w4}). It is also known that
stationary black hole solutions merely represent the asymptotic final
states of gravitational collapses of localized bodies \cite{c1,c2}.
As it follows from perturbation analyses (see e.g.  \cite{ch} for a
detailed study of the related issues) that the time scale characterizing
the convergence to such an asymptotic final state is in inverse ratio
to surface gravity. Thereby these considerations do also manifest
that extreme stationary black holes can be considered as representing
physically inaccessible limits \cite{bch,c2}.

\medskip

Both of the above versions of the third law refer to the behaviour of
thermodynamical systems extrapolated to $T=0$. It is an inherent
nature of such a process that a careful specification of the adequate
representation space is needed.  The principal point of this paper is
to demonstrate that, opposed to the general belief, the extreme black
hole solutions are in certain sense inappropriate limits of
non-extreme ones. Although we agree with the argument of \cite{i1,i2}
that to overcome the underlying difficulties the third law should be
given a precise formulation as a dynamical law no attempt is made here
pointing towards this direction.

\medskip

There are numerous indications that certain physical properties of
stationary black hole spacetimes may change discontinuously at the
extremal limit. Probably the most familiar of these is that while the
black hole region is filled with trapped surfaces in case of
non-extreme black hole spacetimes no trapped surface does exist in
extremal ones.  It is also known that the null geodesic generators of
a future event horizon are geodesically complete in both the past and
future directions, while that of a non-extreme horizon have to be
incomplete in the past direction. In addition, in all extreme cases
the horizon Killing vector field, tangential to the null geodesic
generators of the event horizon, does not vanish at any finite point
of the spacetime.  Moreover, on any of the `t=const' hypersurfaces,
``the horizon is at an infinite distance''. By contrast, in the case
of non-extreme stationary black hole spacetimes the `t=const'
hypersurfaces smoothly extend to the bifurcation surface \cite{rw1}.
Furthermore, this bifurcation surface is at a finite distance on them
and the horizon Killing vector field vanishes there.  Essentially
these differences are reflected by  the main result of \cite{gk}
concluding that the topology of extreme and non-extreme static
Euclidean black holes are intrinsically different.  It is also argued
there that the entropy of all extremal black holes vanishes.

\medskip

Parallel to the above considerable differences, a discontinuous change
in the character of the horizon Killing vector field, $k^a$, also
occurs. It is known that in the non-extreme case, i.e.  whenever
$\kappa\not= 0$, the  squared norm, $k^ak_a$, of $k^a$ has to change
sign across the event horizon, $\mathcal{N}$. It is, in fact, negative
(respectively, positive) at least in a sufficiently small
neighbourhood of the horizon in the domain of outer communication
(respectively, black hole region) side. In other words, $k^a$ must be
timelike outside while spacelike inside the horizon in the relevant
neighbourhood.

The extremal black holes differ significantly. In particular, in case
of an extreme Reissner-Nordstr\"om spacetime $k^a$ must be timelike on
both sides of $\mathcal{N}$.  Indeed, the change of the character of
the horizon Killing vector field in the Kerr case is even more
peculiar. It can be justified that $k^a$ must be timelike on both
sides of $\mathcal{N}$ near the poles (for $\theta\in[0,\pi]$
satisfying $\sin(\theta)<\sqrt{3}-1$), spacelike on both sides near
the equatorial. Moreover, these domains are separated by two
hypersurfaces transverse to $\mathcal{N}$ on which the horizon Killing
vector field is null. It is then not easy to create  a sensible
argument concluding that an extreme Kerr  black hole is essentially
the same as the near extreme ones.

This peculiar behaviour of the horizon Killing vector field was found
to extend over more general Einstein-matter black holes. In \cite{r}
spacetimes with an extreme horizon were considered. Here, as in
\cite{r1}, gravity was assumed to be coupled to either a Klein-Gordon,
Higgs, Maxwell-Yang-Mills or Maxwell-Yang-Mills (-Higgs, -dilaton)
system. It was shown then that whenever the horizon possesses
topologically spherical cross-sections -- wide enough setting to
include both the asymptotically flat and asymptotically anti-de-Sitter
stationary black hole spacetimes -- there must always exist a section
of the horizon so that the horizon Killing vector field is timelike on
both sides in a sufficiently small neighbourhood of the selected
section.

\medskip

On physical grounds, one would expect that the above mentioned
differences between extreme and non-extreme black holes should somehow
be reflected through certain semiclassical effects. In fact, some of
the recent investigations justified this expectation.

It was found, for instance, that the particle production by an object
collapsing into either an extremal Reissner-Nordstr\"om or Kerr black
hole has, in contrast to the non-extreme case, a non-thermal spectrum
\cite{lrs,ro}. In particular, no temperature, and hence no
thermodynamic entropy, can be associated with such a system. Thus an
extremal Reissner-Nordstr\"om or Kerr  black hole cannot be regarded
as the thermodynamic limit of near extremal configurations.

Somewhat complementary is the result covered by \cite{aht}. It was
demonstrated there that if static zero temperature semiclassical black
hole solutions do exist they need to be microscopical and they must not
smoothly join onto the classical extreme Reissner-Nordstr\"om ones.

Both of these investigations support the conclusion that even if there
exist any zero temperature static black holes within the full
semiclassical theory of gravity those spacetimes should be isolated in
the space of solutions from the classical extreme Reissner-Nordstr\"om
black holes.

The above cited classical and semiclassical results suggest that the
conventional way of taking a limit in thermodynamic quantities at
extremal black hole solutions is suspect if not incorrect. It seems to
us that the essential difficulties are due to the fact that in the
traditional formulation of the third law in black hole mechanics only
certain macroscopic parameters -- such as the total mass, angular
momentum and electric charge -- are pro forma adjusted to attain the
extremal states.  However, no empirical evidence or statistical
mechanical considerations is known that could justify the validity of
such extrapolations.  All the above results indicate that a better
understanding of this point is needed to avoid any maladjustment and
to overcome the present puzzling situation to get finally an
appropriate form of the third law of black hole thermodynamics.

\renewcommand{\baselinestretch}{1.12}

\end{document}